\documentclass[onecolumn,authoryear]{els-mrw} 

\usepackage{amsmath,amssymb,amsfonts,amsthm,makeidx,graphicx}
\usepackage{txfonts}
\usepackage{helvet}

\def\DeclareAbbreviation#1#2{%
   \DeclareRobustCommand*#1{\@journalname{#2}}}
\def\@journalname#1{{\normalfont#1}}
\DeclareAbbreviation\aj{AJ}
\DeclareAbbreviation\araa{ARA\&A}
\DeclareAbbreviation\apj{ApJ}
\DeclareAbbreviation\apjl{ApJL}
\DeclareAbbreviation\apjs{ApJS}
\DeclareAbbreviation\ao{Appl.\ Opt.}
\DeclareAbbreviation\apss{Ap\&SS}
\DeclareAbbreviation\aap{A\&A}
\DeclareAbbreviation\aapr{A\&AR}
\DeclareAbbreviation\aaps{A\&AS}
\DeclareAbbreviation\azh{AZh}
\DeclareAbbreviation\baas{BAAS}
\DeclareAbbreviation\jrasc{JRASC}
\DeclareAbbreviation\memras{MmRAS}
\DeclareAbbreviation\mnras{MNRAS}
\DeclareAbbreviation\pra{Phys.\ Rev.\ A}
\DeclareAbbreviation\prb{Phys.\ Rev.\ B}
\DeclareAbbreviation\prc{Phys.\ Rev.\ C}
\DeclareAbbreviation\prd{Phys.\ Rev.\ D}
\DeclareAbbreviation\pre{Phys.\ Rev.\ E}
\DeclareAbbreviation\prl{Phys.\ Rev.\ Lett.}
\DeclareAbbreviation\pasp{PASP}
\DeclareAbbreviation\pasj{PASJ}
\DeclareAbbreviation\qjras{QJRAS}
\DeclareAbbreviation\raa{Res.\ Astron.\ Astronphys.}
\DeclareAbbreviation\skytel{S\&T}
\DeclareAbbreviation\solphys{Sol.\ Phys.}
\DeclareAbbreviation\sovast{Soviet\ Ast.}
\DeclareAbbreviation\ssr{Space\ Sci.\ Rev.}
\DeclareAbbreviation\zap{ZAp}
\DeclareAbbreviation\nat{Nature}
\DeclareAbbreviation\na{New Astron.}
\DeclareAbbreviation\iaucirc{IAU\ Circ.}
\DeclareAbbreviation\aplett{Astrophys.\ Lett.}
\DeclareAbbreviation\apspr{Astrophys.\ Space\ Phys.\ Res.}
\DeclareAbbreviation\bain{Bull.\ Astron.\ Inst.\ Netherlands}
\DeclareAbbreviation\fcp{Fund.\ Cosmic\ Phys.}
\DeclareAbbreviation\gca{Geochim.\ Cosmochim.\ Acta}
\DeclareAbbreviation\grl{Geophys.\ Res.\ Lett.}
\DeclareAbbreviation\jcp{J.\ Chem.\ Phys.}
\DeclareAbbreviation\jgr{J.\ Geophys.\ Res.}
\DeclareAbbreviation\jqsrt{J.\ Quant.\ Spectrosc.\ Radiat.\ Transfer}
\DeclareAbbreviation\memsai{Mem.\ Soc.\ Astron.\ Italiana}
\DeclareAbbreviation\nphysa{Nucl.\ Phys.\ A}
\DeclareAbbreviation\physrep{Phys.\ Rep.}
\DeclareAbbreviation\physscr{Phys.\ Scr.}
\DeclareAbbreviation\planss{Planet.\ Space\ Sci.}
\DeclareAbbreviation\procspie{Proc.\ SPIE}
\DeclareAbbreviation\aip{AIP Conf.\ Proc.}
\DeclareAbbreviation\asp{ASP Conf.\ Ser.}

\begin{document}

\chapter{Superluminous supernovae}\label{chap1}

\author[1,2,3]{Takashi J. Moriya}%

\address[1]{\orgname{National Astronomical Observatory of Japan}, \orgaddress{2-21-1 Osawa, Mitaka, Tokyo 181-8588, Japan}}
\address[2]{\orgname{Graduate Institute for Advanced Studies, SOKENDAI}, \orgaddress{2-21-1 Osawa, Mitaka, Tokyo 181-8588, Japan}}
\address[3]{\orgname{Monash University}, \orgdiv{School of Physics and Astronomy}, \orgaddress{Clayton, Victoria 3800, Australia}}

\articletag{Chapter Article tagline: update of previous edition,, reprint..}

\maketitle

\begin{glossary}[Glossary]
\term{Circumstellar matter} Materials surrounding supernova progenitors. \\
\term{Luminosity function} Peak luminosity distribution of supernovae. \\
\term{Magnetar} Strongly magnetized rapidly rotating neutron star. \\
\term{Type~I SNe} Supernovae without spectroscopic features of hydrogen. Type~Ib SNe show helium features, while Type~Ic SNe show neither hydrogen nor helium features. \\
\term{Type~II SNe} Supernovae with spectroscopic features of hydrogen. \\
\term{Type~IIn/Ibn/Icn SNe} Supernovae with narrow emission features indicating the existence of dense circumstellar matter. \\
\end{glossary}

\begin{glossary}[Nomenclature]
\begin{tabular}{@{}lp{34pc}@{}}
CCSN & Core-collapse Supernova\\
CSM &Circumstellar Matter\\
FBOT &Fast Blue Optical Transient\\
FRB &Fast Radio Burst\\
GRB &Gamma-Ray Burst\\
LSST & Legacy Survey of Space and Time \\
PS1 &Pan-STARRS1\\
PTF &Palomar Transient Factory\\
SLSN &Superluminous Supernova\\
SN &Supernova\\
ZAMS &Zero-Age Main Sequence\\
ZTF &Zwicky Transient Facility\\
\end{tabular}
\end{glossary}

\begin{abstract}[Abstract]
Superluminous supernovae (SLSNe) are a population of supernovae (SNe) whose peak luminosities are much larger than those of canonical SNe. Although SLSNe were simply defined by their peak luminosity at first, it is currently recognized that they show rich spectroscopic diversities including hydrogen-poor (Type~I) and hydrogen-rich (Type~II) subtypes. The exact mechanisms making SLSNe luminous are still not fully understood, but there are mainly four major suggested luminosity sources (radioactive decay of $^{56}$Ni, circumstellar interaction, magnetar spin-down, and fallback accretion). We provide an overview of observational properties of SLSNe and major theoretical models for them. Future transient surveys are expected to discover SLSNe at high redshifts which will provide a critical information in revealing their nature.
\end{abstract}

\begin{BoxTypeA}[chap1:box1]{Key Points}
\begin{itemize}
\item SLSNe are a class of SNe that become more luminous than around $-20~\mathrm{mag}$ in the optical. Broadly, SLSNe have two spectroscopic types: hydrogen-poor (Type~I) and hydrogen-rich (Type~II).
\item Hydrogen-poor (Type~I) SLSNe are characterized by O~\textsc{ii} absorption features around the peak luminosity. Their luminosity sources and progenitors are still debated.
\item Most hydrogen-rich (Type~II) SLSNe have narrow hydrogen emission features (Type~IIn) indicating the existence of dense CSM. Thus, they are powered by the interaction between SN ejecta and dense, massive ($\gtrsim 5~\mathrm{M_\odot}$) CSM. The origin of such a CSM is still not clear.
\end{itemize}
\end{BoxTypeA}

\section{Introduction}\label{chapslsn:introduction}
Superluminous supernovae (SLSNe) are a class of SNe that become more luminous than other kind of SNe at their peak luminosity. The existence of a population of such a luminous SN was not recognized until the 2000s. The first glimpse was found in SN~1999as \citep{1999IAUC.7128....1K,deng2001}, but its nature remained unclear for a long time. SLSNe started to be discovered frequently when unbiased SN surveys were started to be conducted in the 2000s. The first well-observed SLSNe include SN~2005ap \citep{quimby2007}, SN~2006gy \citep[e.g.,][]{smith2008sn06gy,smith2010sn06gy}, and SCP06F6 \citep{barbary2009}. Some mysterious SNe whose origin was unclear when they were discovered were later identified as SLSNe \citep[e.g.,][]{2013ApJ...778..168K,cartier2022}. Several hundred SLSNe have been identified so far \citep{nicholl2021review}. Several review papers on SLSNe are available for further reading \citep{gal-yam2012,gal-yam2019review,howell2017,moriya2018,inserra2019,nicholl2021review}.

\section{Definition}
SLSNe have intrinsically higher luminosity than other SNe. SLSNe are initially defined as SNe that have peak luminosity more luminous than $-21~\mathrm{mag}$ in the optical (see \citealt{gal-yam2012} for an early review). This luminosity cut of $-21~\mathrm{mag}$ is about 10 times larger than the peak luminosities of commonly observed SNe. However, as the number of SN discovery increases, it was recognized that there are many SNe with peak luminosity fainter than $-21~\mathrm{mag}$ showing similar spectroscopic characteristics to SNe with peak luminosity brighter than $-21~\mathrm{mag}$. In other words, it was recognized that SNe that show characteristic spectroscopic features of SLSNe do not necessarily have a clear magnitude cut. For example, the peak luminosity of SNe with hydrogen-free (Type~I) SLSN spectroscopic features can be as faint as around $-20~\mathrm{mag}$ as discussed in Section~\ref{sec:typeilightcurve}. In addition, some rapidly-evolving SNe like so-called fast blue optical transients (FBOTs) exceed $-21~\mathrm{mag}$ in a short time ($\lesssim 3~\mathrm{days}$) after explosion, but they have different spectroscopic features from canonical SLSNe and they are usually not referred to as SLSNe \citep[e.g.,][]{ho2023}. Thus, at least for Type~I SLSNe, SLSNe are defined as a population of luminous SNe with characteristic spectroscopic features rather than SNe that exceed a certain luminosity cut. The observed number of other SLSNe such as Type~II SLSNe (Section~\ref{sec:typeiislsn}) is still too limited to characterize them by their spectroscopic features. In such cases, SNe more luminous than around $-20~\mathrm{mag}$ are naively called SLSNe following the luminosity range of Type~I SLSNe.

\section{Observational properties}
As in the case of other SNe, SLSNe can be broadly classified into two spectroscopic classes based on the presence or absence of hydrogen features in their spectra. Type~I SLSNe are SLSNe without hydrogen features and Type~II SLSNe are SLSNe with hydrogen features. We summarize their observational properties in this section.

\subsection{Type~I SLSNe}
Type~I SLSNe have been actively observed by many transient surveys. Summaries of Type~I SLSN samples from major transient surveys so far can be found in \citet[][PTF]{decia2018}, \citet[][PTF]{quimby2018}, \citet[][PS1]{lunnan2018}, \citet[][DES]{angus2019}, and \citet[][ZTF]{chen2023ztf1}. \citet{gomez2024} also provides a summary of Type~I SLSN properties.

\subsubsection{Spectroscopic properties}
\begin{figure}[t]
\centering
\includegraphics[width=.5\textwidth]{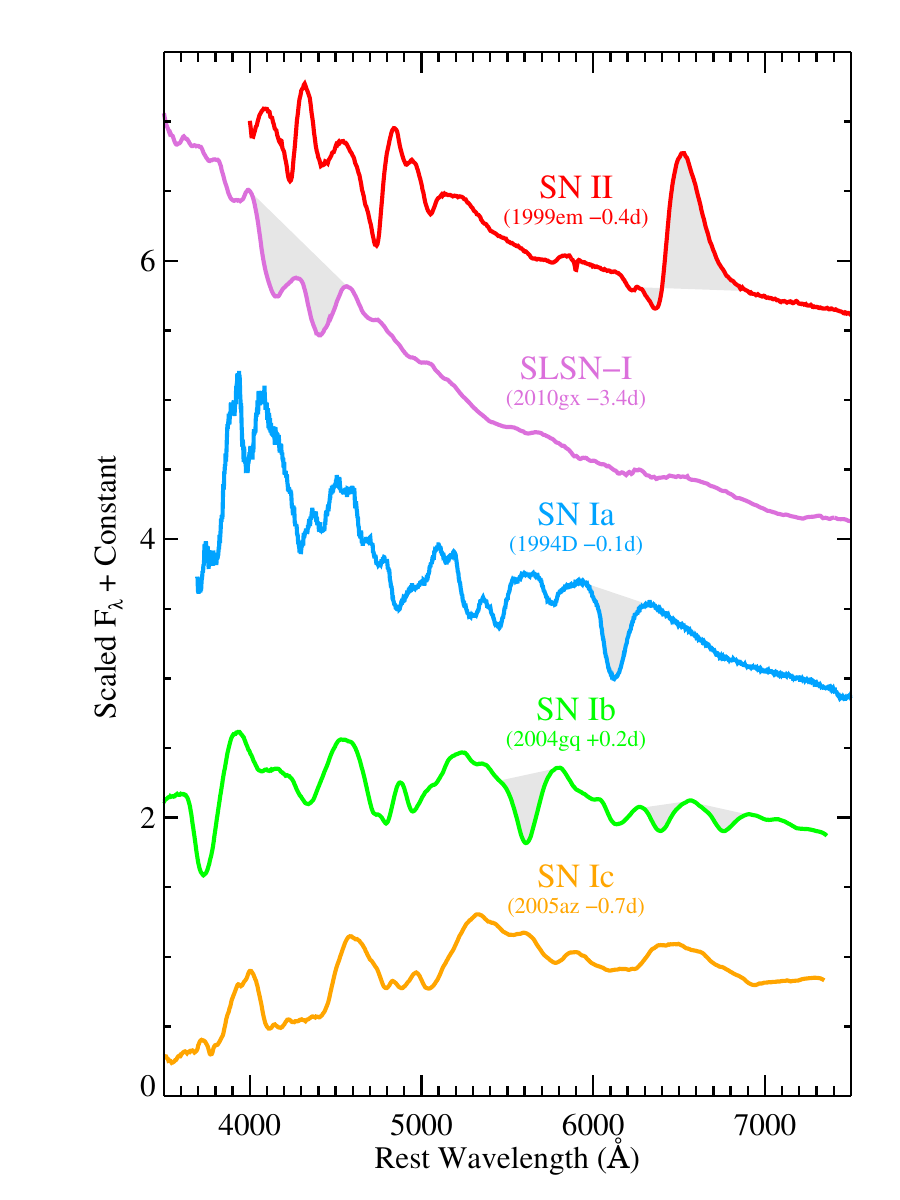}
\caption{
Spectra of Type~I SLSNe compared with other hydrogen-free SNe. The shaded regions indicate the characteristic spectroscopic feature of each SN type.  Type~I SLSNe are characterized by the O~\textsc{ii} absorption features between $3000~\AA$ and $5000~\AA$. The characteristic features of the other SN types are H~\textsc{i} (Type~II), Si~\textsc{ii} (Type~Ia), and He~\textsc{i} (Type~Ib). Reproduced with permission from \citet{quimby2018}.
}
\label{chapslsn:spec}
\end{figure}

Type~I SLSNe are characterized by their O~\textsc{ii} absorption features in the wavelength range between 3000~\AA\ and 5000~\AA\ observed around their luminosity peak (\citealt{quimby2011}, Figure~\ref{chapslsn:spec}). No other prominent spectroscopic features appear in optical spectra around the luminosity peak. Thermal excitation is suggested to be sufficient to excite oxygen to form the O~\textsc{ii} features if the photospheric temperature is around $14,000-16,000~\mathrm{K}$ \citep{saito2024}. Thus, the O~\textsc{ii} features can present diverse temporal evolution and strength depending on the photospheric temperature evolution of Type~I SLSNe \citep{konyves-toth2022}. Most Type~I SLSNe do not show helium features and thus they are Type~Ic SNe. Only a few SLSNe are found to have helium (Type~Ib SN) features so far and their peak luminosity is rather faint (aroung $-20~\mathrm{mag}$ in the optical, \citealt{yan2020}). Photospheric velocities estimated by Fe~\textsc{ii} lines exceeds $10,000~\mathrm{km~s^{-1}}$ at around the luminosity peak \citep{nicholl2015,quimby2018}. The temporal evolution of photospheric velocities is overall found to be similar to those of Type~Ic-BL \citep{liu2017modjaz}.

The rest-frame ultraviolet spectra around the light-curve peak of Type~I SLSNe are diverse \citep{vreeswijk2014,nicholl2017gaia,yan2017uv,yan2018uv}. Ultraviolet spectra below 3000~\AA\ have attenuation due to metal absorption, but the amount of attenuation varies. Strong absorptions in the ultraviolet spectra are likely formed by combinations of the absorptions of several metal lines \citep{howell2013,mazzali2016,quimby2018,gal-yam2019}.  

As the photospheric temperature declines, optical spectra starts to show line features of diverse elements as in other Type~I SNe and they evolve to nebular phases. In some cases (e.g., SN~2007bi, \citealt{gal-yam2009}), strong Ca~\textsc{ii} emissions start to appear on top of the photospheric spectral features. The origin of these early strong Ca~\textsc{ii} emissions is still unclear. Around 10 Type~I SLSNe are observed until the entire ejecta become transparent (so-called the ``nebular'' phase). In the nebular phases, Type~I SLSNe are found to have similar spectral features to those of Type~Ic-BL \citep{nicholl2019neb}. They show strong, broad emission lines of O~\textsc{ii} and Ca~\textsc{ii}, for example. This fact indicates that the physical conditions at the central regions of Type~I SLSNe and Type~Ic-BL SNe are likely similar.

Another notable characteristic of Type~I SLSN is the emergence of hydrogen emission at late phases observed in a few Type~I SLSNe \citep{yan2017lateh}. The late-phase hydrogen emission is likely to indicate the existence of detached hydrogen-rich dense circumstellar matter (CSM) surrounding the progenitors, although it may also originate from hydrogen stripped from the companion star of the progenitors \citep{moriya2015bin}. The late-time emergence of hydrogen emission lines is also observed in a couple of Type~Ibc SNe \citep{milisavljevic2015,chen2018sn2017ens,kuncarayakti2018}. Even if no clear interaction signatures are found, the existence of relatively dense CSM surrounding Type~I SLSNe is sometimes imprinted in their spectra \citep[e.g.,][]{lunnan2018uv}.

\subsubsection{Light-curve properties}\label{sec:typeilightcurve}
\begin{figure}[t]
\centering
\includegraphics[width=.5\textwidth]{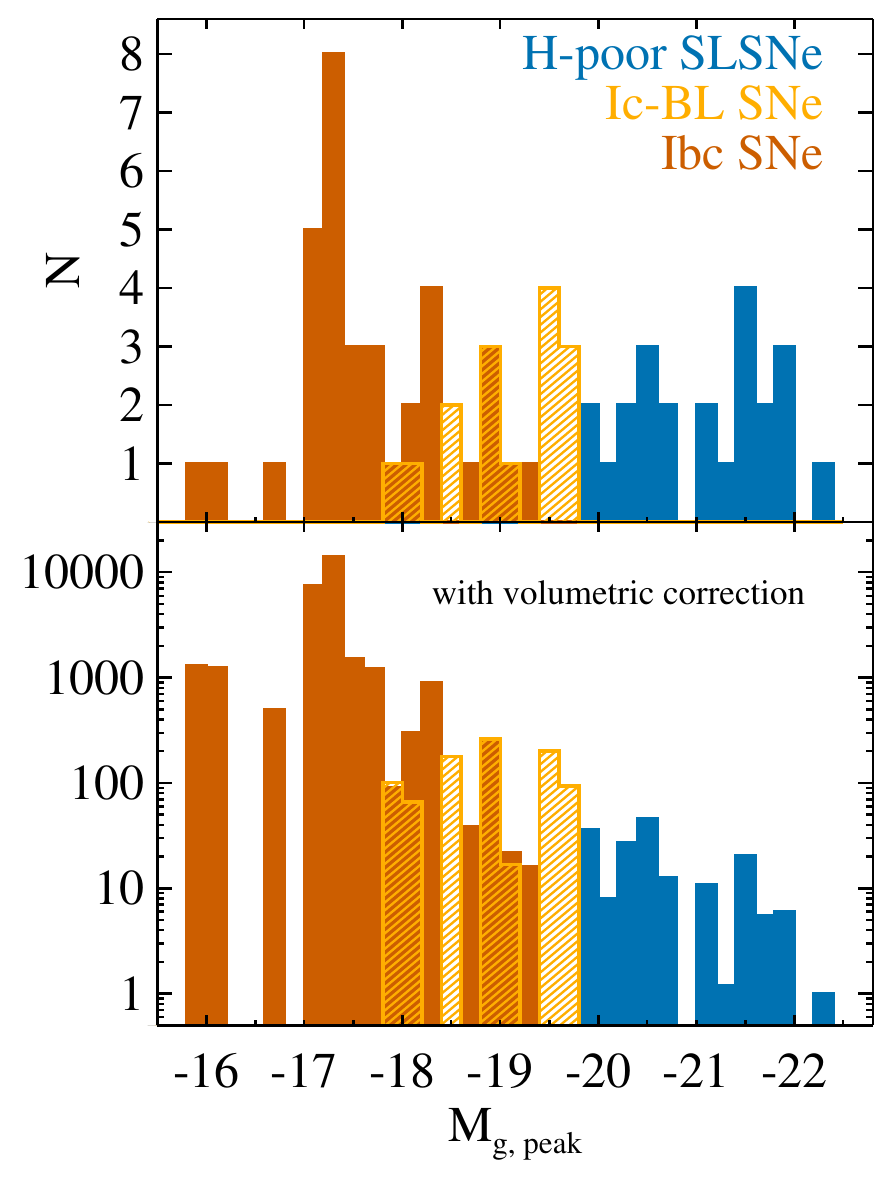}
\caption{Luminosity function of hydrogen-poor SNe. Reproduced with permission from \citet{decia2018}.}
\label{chapslsn:LF}
\end{figure}

The luminosity evolution of Type~I SLSNe shows diversity. Their peak luminosity ranges from $-20~\mathrm{mag}$ to $-23~\mathrm{mag}$ in optical or $3\times 10^{43}~\mathrm{erg~s^{-1}}$ to $8\times 10^{44}~\mathrm{erg~s^{-1}}$ in bolometric (e.g., \citealt{chen2023ztf1}). The peak luminosity distribution of Type~I SLSNe is consistent with an extrapolation from the lower luminosity Type~I SNe including Type~Ibc and Type~Ic-BL SNe (Figure~\ref{chapslsn:LF}). In other words, Type~I SLSNe do not likely make a separate population in hydrogen-free SNe but they are higher luminosity extension of hydrogen-free SNe. Intermediate SNe are briefly discussed in Section~\ref{sec:intermediate}.

The mean rise time of Type~I SLSNe is estimated to be $41.9\pm17.8~\mathrm{days}$ in the \textit{g} band with the $1\sigma$ dispersion in the recent ZTF sample study by \citet{chen2023ztf1}. The rise time is much shorter than the mean rise time of Type~Ibc SNe ($\sim 20~\mathrm{days}$, e.g., \citealt{drout2011}). However, there are Type~I SLSNe with the rise time as short as about 10~days \citep{chen2023ztf1}. To the other extreme, some Type~I SLSNe have a very long rise time exceeding 100~days (e.g., PS1-14bj, \citealt{lunnan2016bj}; SN~2018ibb, \citealt{schulze2024}). These extreme cases are, however, found to be rare. The rise time and the decline time of Type~I SLSNe are positively correlated, i.e., slowly rising Type~I SLSNe tend to decline slowly. There has been a suggestion that Type~I SLSNe can be divided into two populations of slowly evolving Type~I SLSNe and rapidly evolving Type~I SLSNe \citep[e.g.,][]{inserra2018}. It has also been suggested that Type~I SLSNe may have a relation between light-curve decline rate and peak luminosity as in Type~Ia SNe and they could be a potential standardizable candle \citep{inserra2014cos,inserra2021cos,khetan2023nandita}.

The luminosity evolution of Type~I SLSNe is not merely characterized by a simple rise and fall. First, it is known that some Type~I SLSNe show a precursor before the major luminosity increase. The precursor was first identified in SN~2006oz by \citet{leloudas2012}, which showed a precursor luminosity ``bump'' before the major luminosity increase. Subsequently, LSQ14bdq \citep{nicholl2015bdq} and DES14X3taz \citep{smith2016des} are found to have a clear precursor bump. The precursor bump lasts for about 10~days. No spectrum during the bump has been obtained so far, but the color during the bump indicates that the bump should have a very hot spectrum \citep{nicholl2016bump}. Recent studies on a large number of Type~I SLSNe indicate that such a precursor luminosity bump is not a ubiquitous feature of Type~I SLSNe and only a fraction (40\% or less) of Type~I SLSNe show the precursor bump \citep{angus2019}. In one case of SN~2018bsz, the precursor was found to have a gradual increase in luminosity without showing the temporal luminosity decline before the major luminosity increase (\citealt{anderson2018bsz}). In some cases, the early bumps may not be prominent and they may be observed as early flux excess \citep{vreeswijk2017}.

After the luminosity peak, a significant fraction of Type~I SLSNe show undulations in their light curves \citep{inserra2017,chen2023ztf2}. Even in Type~I SLSNe without clear undulations, there may often exist an underlying flux excess in their light curves \citep{chen2017,hosseinzadeh2022}. In some clear cases, we can even observe a secondary luminosity peak in light curves \citep{hosseinzadeh2022}, which are also sometimes observed in Type~Ic SNe that have lower luminosity than Type~I SLSNe \citep[e.g.,][]{gomez2021}. Optical color and spectral line features do not change significantly during the undulations.

Only a couple of Type~I SLSNe have light-curve information beyond 1000~days after explosion \citep[e.g.,][]{blanchard2021}. The light-curve decline rates at the very late phase are found to be diverse even among these cases. The late-phase luminosity evolution might provide an important clue in their luminosity source as discussed in Section~\ref{sec:models}.

\subsubsection{Polarimetric properties}
Polarimetric properties of Type~I SLSNe have been investigated to constrain their ejecta geometry (see \citealt{pursiainen2023} for a summary of Type~I SLSNe with polarimetric observations). No significant polarization is observed in many cases, which indicates that ejecta are not far from spherical symmetry. However, polarization is indeed observed in some cases. For Type~I SLSNe with significant polarization, it is often found that the polarization degree increases with time. This indicates that the asphericity of the ejecta in Type~I SLSNe increases with time. During the period when the significant polarization degree is observed, some signatures of the CSM interaction is also identified in many cases \citep{pursiainen2023}. While most polarimetric observations have been conducted to measure linear polarization, PS17bek is the only Type~I SLSN with circular polarimetric observations \citep{cikota2018}. However, no significant circular polarization is identified in PS17bek.

\subsubsection{X-ray properties}
\begin{figure}[t]
\centering
\includegraphics[width=.425\textwidth]{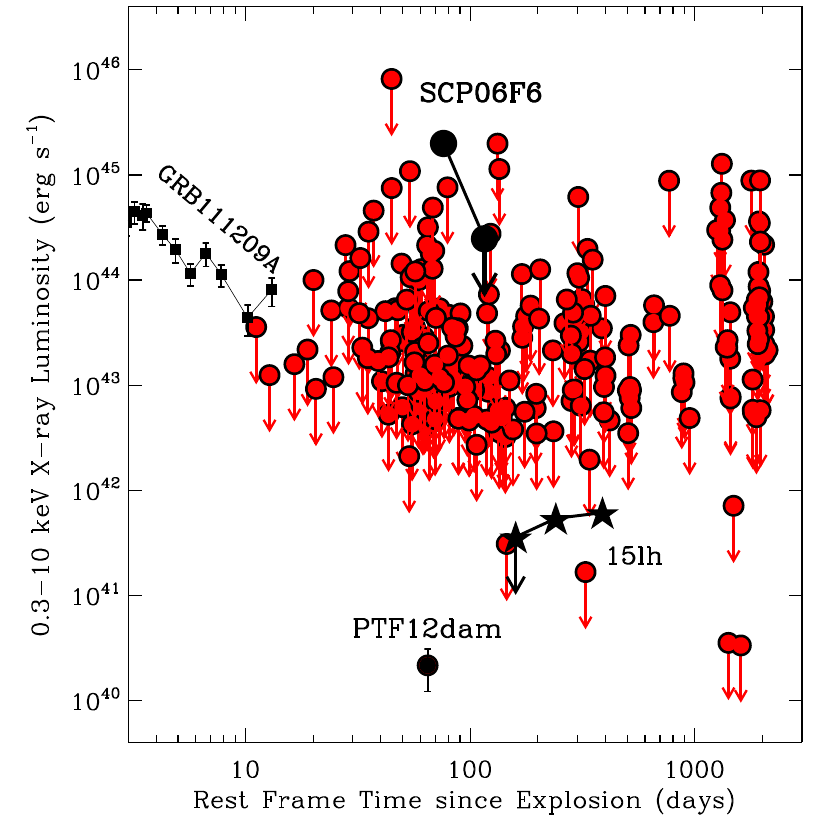}
\includegraphics[width=.4\textwidth]{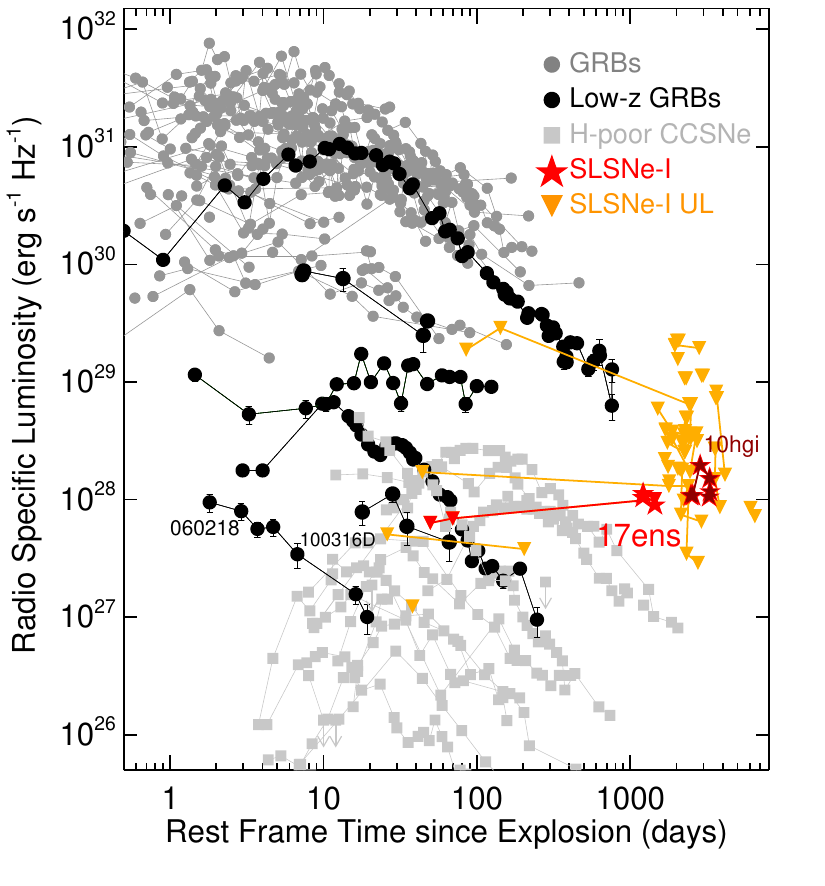}
\caption{Summaries of the observations of Type~I SLSNe in X-rays (left) and radio (right). Reproduced with permission from \citet[][left]{margutti2018} and \citet[][right]{margutti2023}.}
\label{chapslsn:xradio}
\end{figure}

Type~I SLSNe have been extensively observed in X-ray, but they have been detected only in a couple of cases \citep[e.g.,][]{margutti2018}. The first case is SCP06F6 \citep{levan2013}. It was detected in $0.2–2.0~\mathrm{keV}$ by XMM-Newton at around 150~days after the discovery, and its corresponding rest-frame luminosity is $\simeq 10^{45}~\mathrm{erg~s^{-1}}$. This X-ray luminosity is three orders of magnitude higher than those observed in other SNe \citep{levan2013}, making SCP06F6 the most luminous SN observed in X-ray. SCP06F6 was observed again 3 months after the detection, but it was not detected ($<2.5\times 10^{44}~\mathrm{erg~s^{-1}}$, \citealt{levan2013}). Many other Type~I SLSNe were observed in similar epochs with sufficient depths to detect them if they are as bright as SCP06F6, but SCP06F6 remains to be the only detection with such a high luminosity among Type~I SLSNe (\citealt{margutti2018}, Figure~\ref{chapslsn:xradio}).

PTF12dam is another case of the X-ray detection from a Type~I SLSN \citep{margutti2018}. It was detected by the Chandra X-ray Observatory at around the optical luminosity peak, and its X-ray luminosity was $\sim 2\times 10^{40}~\mathrm{erg~s^{-1}}$ in $0.3-10~\mathrm{keV}$. The X-ray luminosity is, however, consistent with the expected diffuse X-ray luminosity from the underlying star forming activities. Thus, it is possible that a significant fraction of the observed X-ray luminosity is not from PTF12dam itself.

Finally, a potential Type~I SLSN ASASSN-15lh \citep{dong2016} was detected in X-ray \citep{margutti2017}, but its nature as a Type~I SLSN has been debated and questioned \citep{leloudas2016}. 

\subsubsection{Mid-infrared properties} 
Some observations of Type~I SLSNe in mid-infrared are currently available. SN~2018bsz, a nearby Type~I SLSN, was observed by Spitzer and it was detected in $3.6~\mathrm{\mu m}$ and $4.5~\mathrm{\mu m}$ at around 400~days and 550~days after the \textit{r} band luminosity peak in the rest frame \citep{chen2021mir}. It was also detected by WISE in $3.4~\mathrm{\mu m}$ and $4.6~\mathrm{\mu m}$ at around 250~days and 400~days after the \textit{r} band luminosity peak. \cite{sun2022} investigated WISE data, and found some additional Type~I SLSNe with mid-infrared detections. The mid-infrared detection indicates the existence of dust in Type~I SLSNe created before and/or after their explosions.

\subsubsection{Radio properties}
Radio information of Type~I SLSNe is suggested to play a key role in constraining their powering mechanism (Section~\ref{sec:magnetar}), and radio follow-up observations of Type~I SLSNe have been conducted actively (see \citealt{margutti2023} for a current summary; Figure~\ref{chapslsn:xradio}). A couple of Type~I SLSNe have been detected in radio. Type~I SLSN PTF10hgi was the first case of the radio detection from Type~I SLSNe (e.g., \citealt{eftekhari2019}). It was detected in $1-20$~GHz from around 7.5~years after the explosion and its radio luminosity was $\simeq 10^{28}~\mathrm{erg~s^{-1}~Hz^{-1}}$ at $6~\mathrm{GHz}$.

SN~2017ens \citep{chen2018sn2017ens} is a SN that reached the peak luminosity of $-21.1~\mathrm{mag}$ in the \textit{g} band and its luminosity is consistent with being a Type~I SLSN. Although it did not show the characteristic O~\textsc{ii} features of Type~I SLSNe, they might have been missed because of infrequent spectroscopic observations. SN~2017ens first had broad features similar to those of Type~Ic-BL at around the peak luminosity. However, it showed Type~IIn SN features after 160~days since the luminosity peak which indicate that the ejecta started to interact with a detached hydrogen-rich dense CSM. SN~2017ens was detected in $3-10~\mathrm{GHz}$ from around 3.3~years after explosion, and its radio luminosity was $\simeq 10^{28}~\mathrm{erg~s^{-1}~Hz^{-1}}$ at $6~\mathrm{GHz}$ \citep{margutti2023}. Given the presence of the late-phase CSM interaction signatures in optical spectra, the radio luminosity may originate from the CSM interaction.

\subsubsection{Gamma-ray properties}
Possible gamma-ray detection from SN~2017egm by the Fermi satellite is reported by \citet{li2024gamma}. The gamma-ray ($500~\mathrm{MeV}-500~\mathrm{GeV}$) was detected at $100-150~\mathrm{days}$ after the discovery.

\subsubsection{Association with gamma-ray bursts}
A potential Type~I SLSN, SN~2011kl, was associated with ultra-long GRB~111209A \citep{greiner2015}. The spectrum of SN~2011kl was not good enough to identify their spectral features, but it reached $-20~\mathrm{mag}$ at the peak luminosity. Long GRB 140506A was associated with a possible luminous blue SN component which might indicate a potential association between normal long GRBs and SLSNe \citep{kann2024}. However, no spectrum was obtained for the potential SN component. Association between GRBs and SLSNe has not been fully investigated and further observational studies are required.

\subsubsection{Host environments}\label{sec:host}
\begin{figure}[t]
\centering
\includegraphics[width=.9\textwidth]{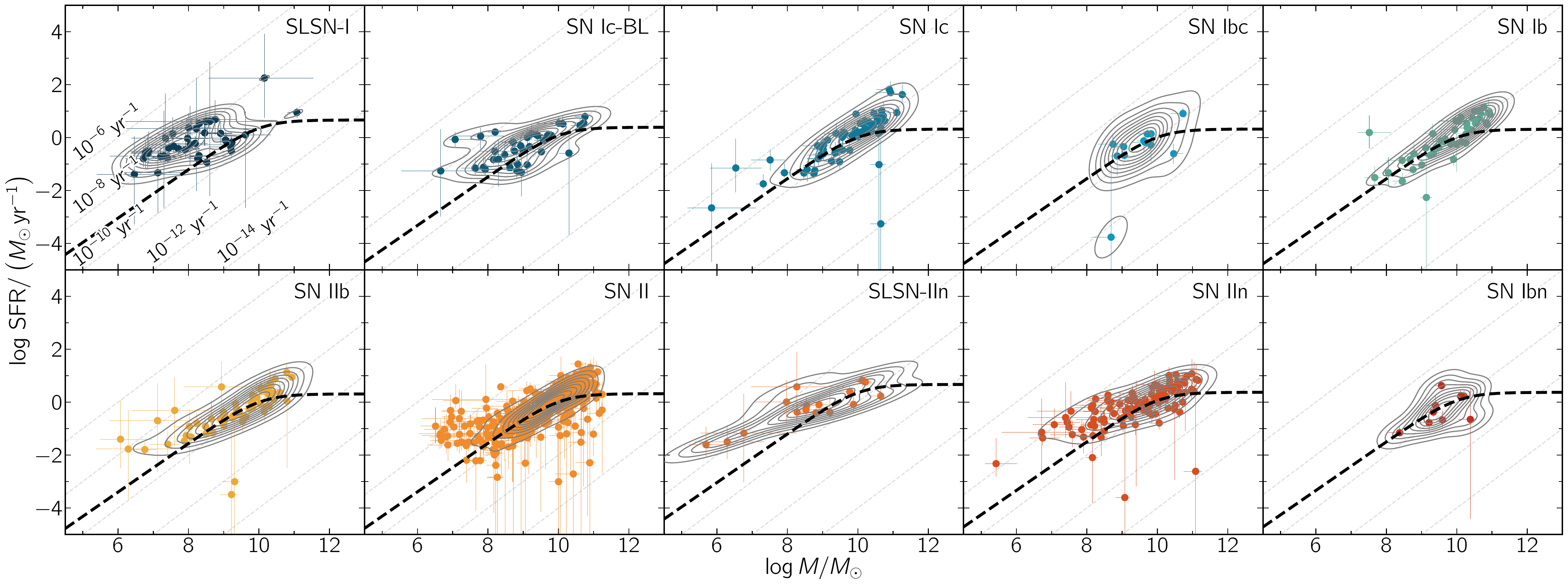}
\caption{
Stellar masses and star-formation rates (SFRs) of SN host galaxies in the PTF core-collapse SN sample. The host galaxies of Type~I SLSNe tend to have high star-formation efficiencies. Reproduced with permission from \citet{schulze2021}.
}
\label{chapslsn:host}
\end{figure}

Type~I SLSNe are observed to prefer low-metallicity environments \citep{neill2011}. Most Type~I SLSNe are observed below around $0.5~\mathrm{Z_\odot}$ \citep{perley2016,chen2017host,schulze2018}, although some SLSNe such as SN~2017egm are exceptionally found in high metallicity environments that are similar to typical core-collapse SNe \citep{bose2018,nicholl2017egm,chen2017egmenv}. The host galaxies of Type~I SLSNe tend to be low stellar-mass galaxies \citep{wiseman2020}. They also tend to have high star-formation efficiencies of around $10^{-9}-10^{-7}~\mathrm{yr^{-1}}$ (e.g., \citealt{schulze2021}; Figure~\ref{chapslsn:host}). This might indicate that Type~I SLSNe are explosions of very massive stars that occur immediately after the star formation, although simply associating high star-formation efficiencies to massive progenitors has been questioned \citep{cleland2023}. Some Type~I SLSNe are also found to be associated with dense molecular clouds with active star formation \citep{arabsalmani2019,hatsukade2020}. Some host galaxies of Type~I SLSNe are interacting galaxies \citep{chen2017,cikota2017,orum2020} or compact dwarf irregular galaxies with extremely strong emission lines likely experiencing active star formations \citep{leloudas2015host,lunnan2015,angus2016}. Type~I SLSNe are found to have a tendency to explode further away from the host galaxy center than any other types of SNe as well as long GRBs \citep{hsu2024}. 

The similarities between the host galaxies of Type~I SLSNe and long GRBs have been explored in many studies \citep[e.g.,][]{lunnan2014}. While some differences among them are pointed out \citep{leloudas2015host,hsu2024}, their host galaxies are rather similar without much statistical differences \citep{japelj2016,taggart2021}. The host galaxies of fast radio bursts (FRBs) are often found to be different from those of Type~I SLSNe, although the host galaxies of repeating FRBs might have some similarities \citep{bhandari2022}.

\subsubsection{Transitional SNe}\label{sec:intermediate}
Type~I SNe with the light-curve peak magnitude range of $\sim -20~\mathrm{mag} - \sim -19~\mathrm{mag}$ in optical show diverse spectroscopic features \citep{gomez2022}. Some of them show hot spectra similar to Type~I SLSNe while others show cool spectra with more absorption features similar to Type~I SNe. Some Type~I SNe in this luminosity range have relatively slow photospheric velocities than those of Type~I SLSNe and Type~I SNe that may indicate an existence of interesting intermediate population in this luminosity range \citep[e.g.,][]{blanchard2019int}.

\subsection{Type~IIn and Type~II SLSNe}\label{sec:typeiislsn}
While more than 100 hydrogen-poor (Type~I) SLSNe have been found so far, the number of observed hydrogen-rich SLSNe is still far less (of the order of 10). Therefore, the spectroscopic features that characterize hydrogen-rich SLSNe have not been carefully investigated, and hydrogen-rich SLSNe are still classified based on their peak luminosities. Following the luminosity range of Type~I SLSNe, SNe with hydrogen features having the peak magnitude brighter than around $-20$~mag in optical are often referred as hydrogen-rich (Type~IIn or Type~II) SLSNe.

Most hydrogen-rich SLSNe show narrow hydrogen lines in their spectra and they are called Type~IIn SNe. The prototype and the most studied SN of this class is SN~2006gy  \citep[e.g.,][]{smith2010sn06gy}. The luminosity and luminosity evolution of Type~IIn SNe are diverse. The most luminous Type~IIn SLSNe reaches to around $-22.5~\mathrm{mag}$ in optical (e.g., SN~2008fz, \citealt{drake2010sn08fz}; SN~2016aps, \citealt{nicholl2020}) and these bright Type~IIn SLSNe tend to have round light-curve shapes around the luminosity peak. As the peak luminosity of Type~IIn SLSNe becomes smaller, their light-curve evolution tends to decline with a power law (e.g., SN~2010jl, \citealt{fransson2014}). However, some Type~IIn SLSNe are known to evolve quite fast (e.g., SN~2003ma, \citealt{rest2011}), while others evolve very slowly (e.g., SN~2015da, \citealt{tartaglia2020}) regardless of their high luminosity. It is not clear if there is a separate population of Type~IIn SLSNe or they consist of the most luminous end of a continuous Type~IIn SN population \citep{richardson2014}. Some Type~IIn SLSNe emit more than $5\times 10^{51}~\mathrm{erg}$ and the explosion inside such events are clearly distinct from other Type~IIn SNe \citep{nicholl2020}.

Some very luminous transients having spectra that are consistent with Type~IIn SNe appear near the center of AGNs \citep{drake2011,kankare2017}. Because spectra of some AGNs and Type~IIn SNe are known to have similar narrow emission features, it is sometimes difficult to distinguish if they are Type~IIn SLSNe originating from stellar explosions or certain activities of AGNs \citep[e.g.,][]{moriya2017}. Some tidal disruption events may also be confused with Type~IIn SLSNe \citep{blanchard2017,petrushevska2023}. We note that SN~2006gy appeared in an X-ray bright galaxy (NGC~1260, \citealt{wang2016}) which might be an AGN and it may have been identified as a nuclear transient if it appeared at high redshifts.

Type~IIn SLSNe have been often observed in infrared wavelengths and they are found to have observational features of dusts \citep{sun2022}. For example, SN~2006gy was bright in infrared for a long time \citep{miller2010}. SN~2010jl is also well observed in infrared with dust signatures \citep{maeda2013}. 

Not all hydrogen-rich SLSNe are Type~IIn SLSNe. Several hydrogen-rich SLSNe without narrow hydrogen emission lines are known and they are referred as Type~II SLSNe \citep{inserra2018typeii,kangas2022typeii}. SN~2008es is the first SLSN that was identified as a Type~II SLSN \citep{gezari2009,miller2009sn08es}. The number of observed Type~II SLSNe is still small \citep{inserra2018typeii,kangas2022typeii} and the characteristic properties of Type~II SLSNe that distinguish them from less luminous Type~II SNe are still not fully understood.

Host environments of Type~IIn SLSNe are more diverse than those of Type~I SLSNe. They can appear in higher metallicity environments than Type~I SLSNe \citep{neill2011}. Their host galaxies can have higher metallicities and larger stellar masses than Type~I SLSNe as well. They might tend to originate from lower metallicity and lower luminosity galaxies than typical core-collapse SNe \citep{schulze2021}. The host environments of Type~II SLSNe are found to be not so different from those of Type~IIn SLSNe \citep{kangas2022typeii}, but the number of Type~II SLSNe is still too small to make a proper comparison. Some studies suggest that Type~II SLSN environments are similar to those of Type~I SLSNe \citep{inserra2018typeii}.

\begin{figure}[t]
\centering
\includegraphics[width=.7\textwidth]{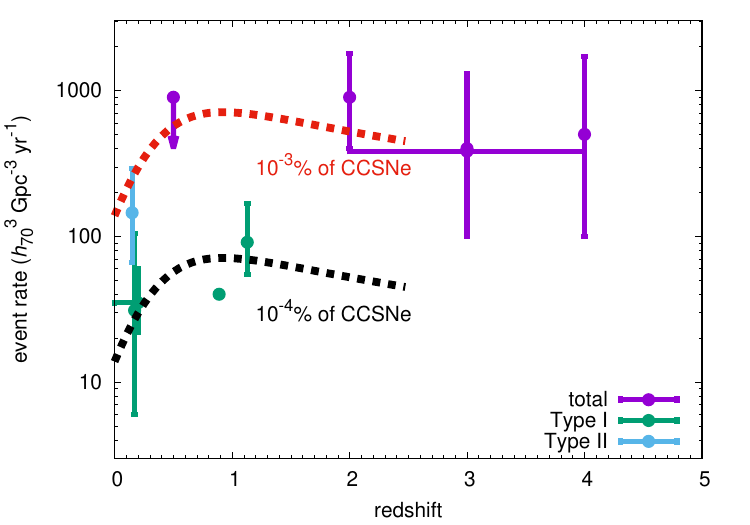}
\caption{
Summary of the estimated event rates of SLSNe. The fractions to core-collapse SNe (CCSNe) are based on the core-collapse SN rate estimated by \citet{strolger2015}.
}
\label{chapslsn:rate}
\end{figure}

\section{Event rates}
The current event rate estimates for SLSNe are summarized in Figure~\ref{chapslsn:rate}.
The event rates of SLSNe were first studied by \citet{quimby2013}. Based on SLSNe discovered by ROTSE-IIIb, they estimated the Type~I SLSN rate at $z\simeq 0.17$ as $32^{+77}_{-26}h_{71}^3~\mathrm{Gpc^{-3}~yr^{-1}}$ and the hydrogen-rich SLSN rate as $151^{+151}_{-82}h_{71}^3~\mathrm{Gpc^{-3}~yr^{-1}}$ at $z\simeq 0.15$. The hydrogen-rich SLSN rate from \citet{quimby2013} includes both Type~IIn and Type~II SLSNe discovered by ROTSE-IIIb. \citet{mccrum2015} estimated the Type~I SLSN rate at $0.3\lesssim z \lesssim 1.4$ as $(3-8)\times 10^{-3}$\% of the core-collapse SN rate. \citet{prajs2017} estimated the Type~I SLSN rate at $z\simeq 1.13$ as $91^{+76}_{-36}h_{70}^3~\mathrm{Gpc^{-3}~yr^{-1}}$ based on the Type~I SLSN sample from Supernova Legacy Survey. \citet{frohmaier2021} estimated Type~I SLSN rate at $z\lesssim 0.2$ as $35^{+25}_{-13}h_{70}^3~\mathrm{Gpc^{-3}~yr^{-1}}$ based on the PTF SLSN sample. Using the public Type~I SLSN data from PS1, \cite{zhao2021} estimated the Type~I SLSN event rate of $40 h_{70}^3~\mathrm{Gpc^{-3}~yr^{-1}}$ at $z\simeq 0.89$ with an unknown error. No updated event rates for Type~IIn and Type~II SLSNe have been obtained after \citet{quimby2013}.

There are some estimates for the total event rates of SLSNe based on photometric samples of SLSNe. For example, based on the fact that no SLSNe were discovered during the Supernova Diversity and Rate Evolution (SUDARE) survey, the total SLSN rate was constrained to be less than $900h_{70}^3~\mathrm{Gpc^{-3}~yr^{-1}}$ at $z\simeq 0.5$ \citep{cappellaro2015}. \citet{cooke2012} estimated the SLSN rate at $z\simeq 2-4$ to be $\sim 400h_{71}^3~\mathrm{Gpc^{-3}~yr^{-1}}$ using two SLSNe at $z=2.05$ and $3.90$ discovered by the photometric SLSN search in the CFHT archival data. The high-redshift SN survey with Subaru/Hyper Suprime-Cam (HSC) provided the SLSN event rates of $\sim 900^{+900}_{-500}h_{70}^3~\mathrm{Gpc^{-3}~yr^{-1}}$ at $z\simeq 2$, $\sim 400^{+900}_{-300}h_{70}^3~\mathrm{Gpc^{-3}~yr^{-1}}$ at $z\simeq 3$, and $\sim 500^{+1200}_{-400}h_{70}^3~\mathrm{Gpc^{-3}~yr^{-1}}$ at $z\simeq 4$ \citep{moriya2019hsc,curtin2019}.

\section{Luminosity sources and progenitors}\label{sec:models}
We provide a brief overview of possible luminosity sources that make SLSNe so bright. For more details of the major proposed luminosity sources, we refer to \citet{moriya2018}. For each possible luminosity source, we also discuss possible progenitors that can realize the conditions required to have the luminosity source.

\subsection{Radioactive decay}\label{sec:radioactive}
The energy released by the radioactive decay of $^{56}\mathrm{Ni}$ synthesized during the SN explosion is a standard powering mechanism of SNe \citep{arnett1996}. Especially, the early luminosity peak of stripped-envelope SNe is mainly powered by the $^{56}\mathrm{Ni}$ decay \citep{bersten2017}. A simple approach extending the standard SN powering mechanism to SLSNe is to have a larger amount of $^{56}$Ni from the explosive nucleosynthesis. The more $^{56}\mathrm{Ni}$ is synthesized, the more radioactive energy is available to make SNe brighter. The synthesized $^{56}\mathrm{Ni}$ decays to $^{56}\mathrm{Co}$ with a decay time of $8.76\pm0.01~\mathrm{days}$ \citep{dacruz1992} and then $^{56}\mathrm{Co}$ decays to $^{56}\mathrm{Fe}$ with a decay time of $111.42\pm0.04~\mathrm{days}$ \citep{funck1992}. The total available energy from this radioactive decay is
\begin{align}\label{eq:ni}
    L_{\mathrm{^{56}Ni-decay}}=\left[6.48\exp\left(-\frac{t}{8.76~\mathrm{days}}\right)+1.44\exp\left(-\frac{t}{111.42~\mathrm{days}}\right)\right]\frac{M_{\mathrm{^{56}Ni}}}{\mathrm{M_\odot}}10^{43}~\mathrm{erg~s^{-1}},
\end{align}
where $M_{\mathrm{^{56}Ni}}$ is the mass of $\mathrm{^{56}Ni}$ synthesized at the explosion (\citealt{nadyozhin1994} with the updated physical values from \url{http://www.nndc.bnl.gov/chart}). The energy from the nuclear decay is mostly released as gamma-rays having the energy of the order of MeV. Because these gamma-rays may not necessarily be absorbed by the SN ejecta, the actual available energy to power the SN luminosity is less than the total decay energy in Equation~(\ref{eq:ni}). 

In order to have a rough estimate for the amount of $^{56}\mathrm{Ni}$ that is required to account for the luminosity of SLSNe, we can use the rise time and peak bolometric luminosity of SLSNe. \citet{arnett1982} analytically showed that the peak luminosity of a SN is the same as the luminosity input at the time of the peak luminosity (so-called "Arnett-law"). If we take the average rise time of $41.9~\mathrm{days}$ for Type~I SLSNe, for example, the $\mathrm{^{56}Ni}$ mass required to explain their peak luminosities above $3\times 10^{43}~\mathrm{erg~s^{-1}}$ is $M_{\mathrm{^{56}Ni}}\gtrsim 3~\mathrm{M_\odot}$ (Equation~\ref{eq:ni}). This $\mathrm{^{56}Ni}$ mass is much higher than the $\mathrm{^{56}Ni}$ mass estimated for typical SNe ($M_{\mathrm{^{56}Ni}}\lesssim 0.3~\mathrm{M_\odot}$, e.g., \citealt{anderson2019}). Even broad-lined Type~Ic SNe that are among the most energetic core-collapse SNe are mostly estimated to have $M_{\mathrm{^{56}Ni}}\lesssim 1~\mathrm{M_\odot}$ \citep{taddia2019lcbl}.

Synthesizing $M_{\mathrm{^{56}Ni}}\gtrsim 3~\mathrm{M_\odot}$ is a challenge in the standard core-collapse SN explosion models. The maximum amount of $\mathrm{^{56}Ni}$ that can be synthesized by core-collapse SNe is estimated to be around $10~\mathrm{M_\odot}$, which requires the explosion energy of $10^{53}~\mathrm{erg}$ \citep{umeda2008}. Some relatively low-luminosity SLSNe that require $M_{\mathrm{^{56}Ni}}\lesssim 10~\mathrm{M_\odot}$ could be consistent with such energetic core-collapse SNe \citep{moriya2019sn07bi,mazzali2019sn07bi}. However, a significant fraction of SLSNe require $M_{\mathrm{^{56}Ni}}\gtrsim 10~\mathrm{M_\odot}$ to explain their peak luminosity. However, pair-instability SNe (PISNe, \citealt{barkat1967,rakavy1967}) are predicted to synthesize up to around $70~\mathrm{M_\odot}$ of $\mathrm{^{56}Ni}$ \citep{heger2002}. PISNe are predicted explosions of very massive stars with the helium core mass between $\sim 65~\mathrm{M_\odot}$ and $\sim 135~\mathrm{M_\odot}$, although the exact mass range depends on uncertainties in, e.g., nuclear reaction rates \citep{takahashi2018,farmer2019,farmer2020}. This helium core mass corresponds to the zero-age main sequence (ZAMS) mass between $\sim 140~\mathrm{M_\odot}$ and $\sim260~\mathrm{M_\odot}$ when mass loss and rotation are ignored \citep{heger2002}. If the rotation is significant, the ZAMS mass of PISN progenitors can be as low as $65~\mathrm{M_\odot}$ through chemically homogeneous evolution \citep[e.g.,][]{chatzopoulos2012pisn}. Stellar mergers are also an important path to form massive stars ending up with PISNe \citep{vigna-gomez2019}.

Because it is required to sustain massive cores until the instability is triggered, PISNe are not expected to occur frequently above a certain metallicity because of strong mass loss at high metallicity \citep{langer2007}. This is consistent with the fact that Type~I SLSNe prefer low metallicity environments. However, the exact metallicity cut is uncertain \citep{sabhahit2023}, and it is also possible to suppress mass loss in high metallicity environments \citep{georgy2017}. The predicted PISN event rates in the local Universe are also lower than the event rates of Type~I SLSNe \citep{briel2022}.

Predicted PISN light curves evolve slowly \citep[e.g.,][]{kasen2011,dessart2013} and they are often consistent with those of slowly evolving SLSNe. In many rapidly evolving SLSNe, the required amount of $\mathrm{^{56}Ni}$ to explain their luminosity becomes more than the estimated ejecta mass \citep{kasen2017}. Therefore, rapidly evolving SLSNe are generally not likely powered by the radioactive decay of $\mathrm{^{56}Ni}$, although mixing of synthesized $\mathrm{^{56}Ni}$ in the ejecta may make luminosity evolution of PISNe faster \citep{kozyreva2015}. However, no significant mixing is predicted in multi-dimensional simulations of PISNe \citep[e.g.,][]{joggerst2011,chen2014pisnmultid}. The nuclear decay of $\mathrm{^{56}Ni}$ as a possible power source of SLSNe is mainly discussed for slowly evolving SLSNe whose light-curve decline rates are consistent with the decay rate of $\mathrm{^{56}Ni}\rightarrow \mathrm{{}^{56}Co}\rightarrow \mathrm{{}^{56}Fe}$. Even if the light-curve evolution is consistent with PISNe, their spectral features are often found to be inconsistent with those predicted by PISNe \citep{dessart2012,dessart2013,jerkstrand2016,jerkstrand2017}. Especially, a large amount of $\mathrm{^{56}Fe}$ is expected to exist in late phases from the nuclear decay of $\mathrm{^{56}Ni}$, but no strong  $\mathrm{^{56}Fe}$ absorption and emission are usually observed in SLSNe. Currently, only SN~2018ibb is found to match most of the predicted PISN properties, but their late-phase spectra have blue flux excess which might be caused by the additional CSM interaction \citep{schulze2024}. 

It is difficult to explain the precursors and light-curve bumps and undulations occurring in different timescales solely by the nuclear decay energy input that is govern by the nuclear decay timescale. Thus, additional luminosity source such as CSM interaction is required to explain the whole luminosity evolution of SLSNe.

\begin{figure}[t]
\centering
\includegraphics[width=.9\textwidth]{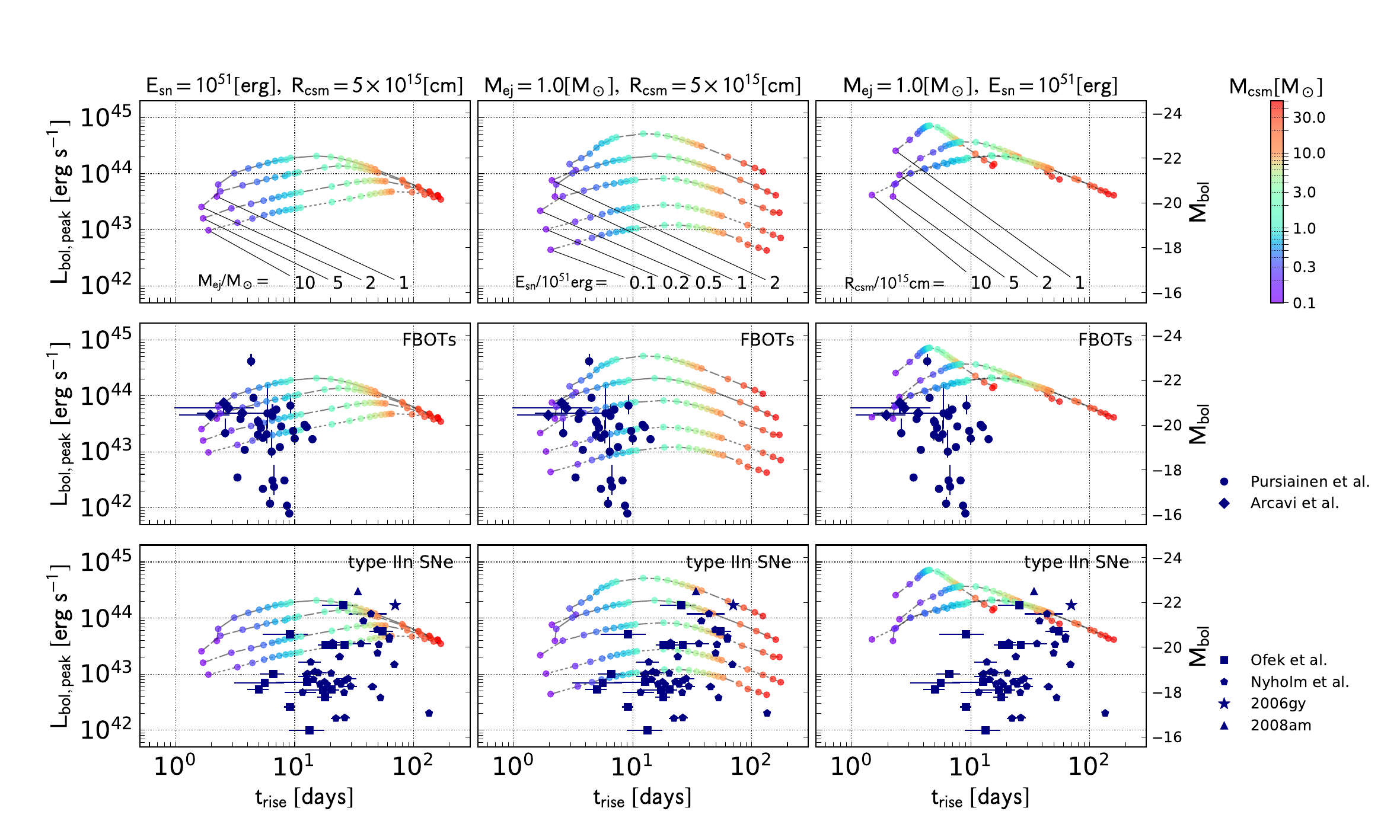}
\caption{Predicted rise times and peak luminosities of interaction-powered SNe and their comparisons with FBOTs and Type~IIn SNe including Type~IIn SLSNe. Reproduced with permission from \citet{suzuki2020}.}
\label{chapslsn:CSM}
\end{figure}

\subsection{Circumstellar interaction}\label{sec:interaction}
The collision of SN ejecta into their CSM can efficiently convert the kinetic energy of the ejecta to radiation especially when the CSM has a comparable mass to the SN ejecta. When the CSM density is high enough, the collision forms a strong radiative shock. The emission is first mostly in X-rays but they can be immediately absorbed through free-free absorption when the CSM density is high. Then the post-shock region can be around $10^4~\mathrm{K}$ to emit photons in optical \citep[e.g.,][]{moriya2013sn06gy}. The unshocked CSM can be optically thick and photons will diffuse out in the dense CSM. This diffusion process can make light curves of interaction-powered SNe broad. The diffusion time can vary depending on the CSM density and radius. Therefore, the CSM interaction model can explain SLSNe of various timescales.

Type~IIn SNe show clear signatures of the CSM interaction in their spectra. Thus, Type~IIn SLSNe are considered to be powered by the CSM interaction as in the case of the lower luminosity Type~IIn SNe \citep[e.g.,][]{moriya2013sn06gy}. Compared to low-luminosity Type~IIn SNe, Type~IIn SLSNe are estimated to have higher explosion energies or more massive CSM because of their higher luminosity. The total radiated energy of Type~IIn SLSNe is of the order of $10^{51}~\mathrm{erg}$, which is a typical SN explosion energy. Thus, if we can efficiently convert the kinetic energy of SN ejecta to radiation through the CSM interaction, it is possible to explain the huge luminosities observed in Type~IIn SLSNe. The total CSM mass around Type~IIn SLSNe are estimated to be $\gtrsim 5~\mathrm{M_\odot}$ (\citealt{chatzopoulos2013,suzuki2020}, Figure~\ref{chapslsn:CSM}). We note that the most luminous Type~IIn SLSNe emit more than $10^{52}~\mathrm{erg}$ in total and the inner explosion sometimes needs to be energetic \citep{nicholl2020,suzuki2021}.

Because observational signatures of Type~IIn SLSNe are dominated by the CSM interaction, it is difficult to identify the nature of the explosions inside. The progenitors of Type~IIn SNe are known to be massive ($\gtrsim 50~\mathrm{M_\odot}$) luminous blue variables (LBVs, \citealt{gal-yam2009iinpro}) and low-mass ($\simeq 10~\mathrm{M_\odot}$) red supergiants \citep{prieto2008}. Among them, some mass eruptions from LBVs such as the Great Eruption of $\eta$ Carinae \citep{humphreys1994} are known to form CSM as massive as $10~\mathrm{M_\odot}$ \citep{morris1999}, although their mass-loss mechanisms are not well understood. If LBVs explode immediately after they experience such an eruptive mass loss, they could be observed as Type~IIn SLSNe. LBVs are considered to be massive stars with ZAMS masses above around $40~\mathrm{M_\odot}$ \citep{humphreys1994}, but they may also originate from stellar mergers of less massive stars \citep{justham2014}. Even if massive stars are not in the LBV phase, they may experience strong mass loss triggered by the strong convection at the innermost layers of massive stars (\citealt{quataert2012,mcley2014}, but see also \citealt{fuller2017}). Phase transition from nuclear matter to the quark-gluon plasma at the center of massive stars after the core collapse is also suggested to result in Type~IIn SLSNe \citep{fischer2018}.

Another mechanism to form massive CSM is common-envelope mass ejection \citep{chevailer2012,ercolino2024}. Unstable mass transfer in binary systems with massive stars can lead to a common-envelope phase. Although the exact outcome of the common-envelope phase is uncertain, one possible consequence is ejection of the massive stellar envelope. If the common-envelope mass ejection occurs shortly before the explosion of massive stars, the SN ejecta can collide with the massive ejected envelope and can be observed as Type~IIn SLSNe.

Extensive mass loss from massive stars can also be related to pulsational pair-instability \citep{woosley2007}. Massive stars slightly below the mass range of PISNe (Section~\ref{sec:radioactive}) can still become dynamically unstable to eject a part of their mass to form a massive CSM. Several mass ejection triggered by this instability can occur sequentially and ejected shells can collide to each other to make Type~IIn SLSNe \citep{woosley2007}. This kind of trainsients triggered by the pulsational pair-instability is called pulsational pair-instability SNe (PPISNe). It is also possible that a core-collapse SN is followed by the pulsational pair-instability. In this case, the SN ejecta can collide to the dense CSM formed by the pulsational pair-instability to become Type~IIn SLSNe. As in the case of PISN progenitors, PPISN progenitors can originate from stellar mergers \citep{vigna-gomez2019}.

It has also been suggested that some Type~IIn SLSNe are related to explosions of white dwarfs (Type~Ia SNe). The late-phase spectrum of SN~2006gy was suggested to be similar to that of Type~Ia SN and its luminosity evolution is also suggested to be consistent with the explosion of a Type~Ia SN within the hydrogen-rich CSM having the mass of $\simeq 10~\mathrm{M_\odot}$ \citep{jerkstrand2020}. A major question for this scenario is how to form such a massive hydrogen-rich CSM around Type~Ia SN progenitors. Some Type~Ia SNe are known to show signatures of hydrogen-rich CSM (so-called Type~Ia-CSM), but their CSM mass is estimated to be of the order of $0.1~\mathrm{M_\odot}$ \citep{sharma2023}. A rare evolutionary path involving common-envelope mass ejection shortly before Type~Ia SN explosions may be able to realize such a massive CSM around Type~Ia SNe \citep{ablimit2021}.

We have discussed Type~IIn SLSNe so far. Although the other types of SLSNe do not show clear signatures of the CSM interaction in their spectra in early phases, the CSM interaction has been considered to be their possible luminosity source. For example, the light curves of Type~I SLSNe can be reproduced by the interaction between SN ejecta and massive hydrogen-poor CSM \citep{sorokina2016}. The precursor bump observed in Type~I SLSNe may also be explained by the existence of a massive CSM \citep{moriya2012bump}. Such a massive hydrogen-poor CSM can be formed by the pulsational pair-instability of massive hydrogen-poor stars \citep{woosley2017} and such massive hydrogen-poor stars can be formed through the chemically homogeneous evolution, for example \citep{marchant2016,mandel2016}. Several Type~I SLSNe have been suggested to be hydrogen-poor PPISNe \citep[e.g.,][]{tolstov2017}. If Type~I SLSNe are mainly powered by the CSM interaction, a major remaining question is why we do not see clear interaction signatures in their spectra. There are hydrogen-poor SNe that show clear CSM interaction signatures in their spectra known as Type~Ibn \citep[e.g.,][]{pastorello2007,hosseinzadeh2017} and Type~Icn SNe \citep[e.g.,][]{fraser2021,gal-yam2022,perley2022}. Thus, we may expect to have similar spectroscopic signatures if Type~I SLSNe are mainly powered by the CSM interaction, although the required CSM mass and density for Type~I SLSNe are expected to be higher. More theoretical investigations on the expected spectroscopic features of hydrogen-poor interaction-powered SNe are required.

Even if the major luminosity source of Type~I and Type~II SLSNe is not the CSM interaction, it is very likely that their properties are often partially affected by the CSM interaction. The luminosity undulations observed in Type~I SLSNe can be explained as an additional effect caused by the CSM interaction. Similarly multiple luminosity bumps in Type~I SLSNe could be related to the existence of the multiple dense CSM components \citep{lin2023}. Because many Type~I SLSNe show the CSM interaction signatures at late phases, it is possible that some effects of the CSM interaction starts to appear in earlier epochs after the major luminosity source ends providing their energy.

\begin{figure}[t]
\centering
\includegraphics[width=.7\textwidth]{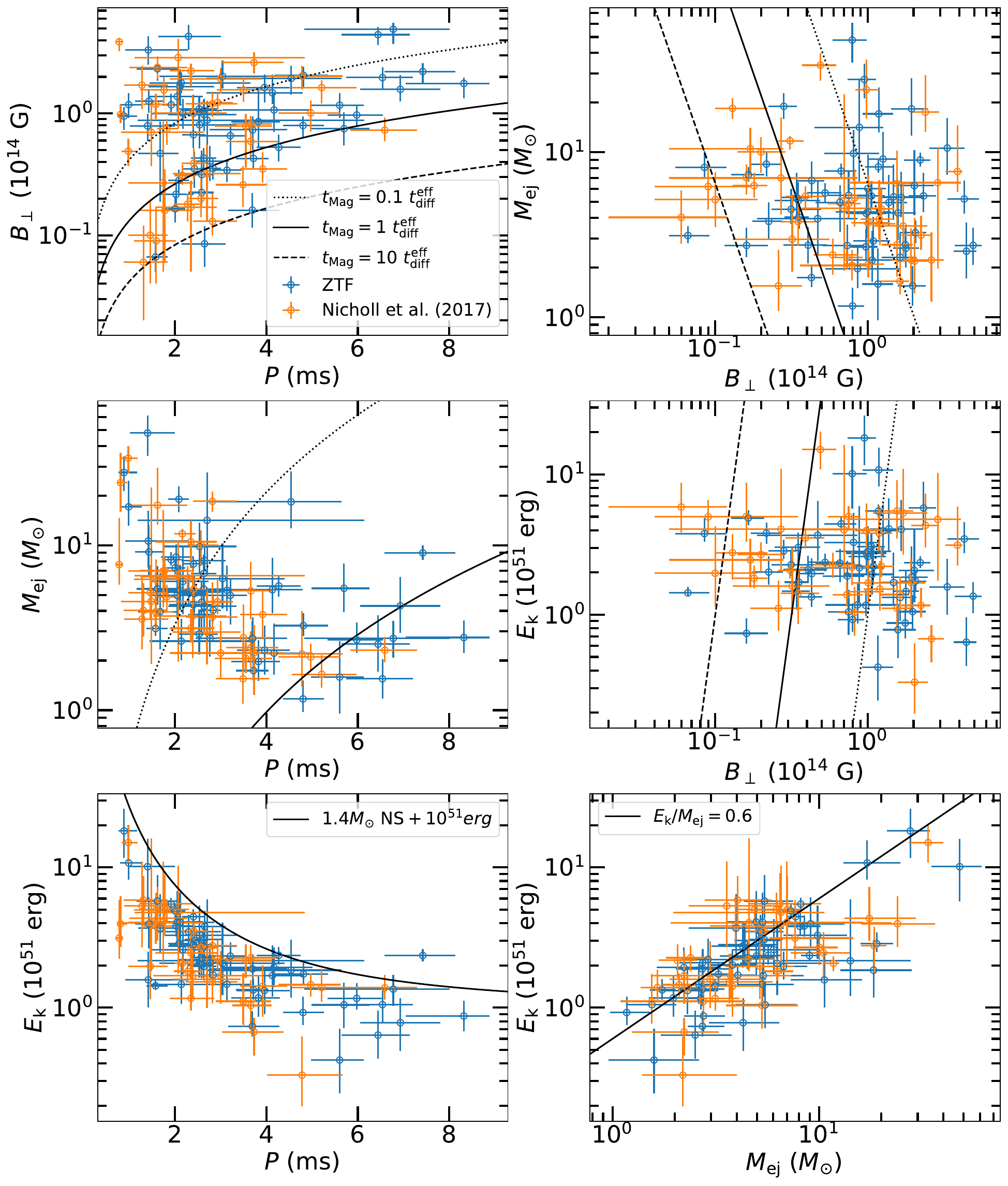}
\caption{Estimated parameters for Type~I SLSNe in the magnetar-powered model. Reproduced with permission from \citet{chen2023ztf2}.}
\label{chapslsn:magnetar}
\end{figure}

\subsection{Spin down of strongly magnetized neutron stars (``magnetars'')}\label{sec:magnetar}
A neutron star may remain at the center of an exploding massive star after a core-collapse SN explosion. If the neutron star has rotation and poloidal magnetic fields (in other word, if the neutron star becomes a ``pulsar''), the rotational energy should be converted to the electromagnetic field mostly as Poynting-flux dominated outﬂows \citep[e.g.,][]{1969ApJ...157..869G}. The total spin-down energy that can be released in this form can be expressed as
\begin{align}\label{eq:mag}
  L_\mathrm{spin-down}=\frac{(l-1)E_p}{t_p}\left(1+\frac{t}{t_p}\right)^{-l},
\end{align}
where $E_p$ is the initial rotational energy of the neutron star and $t_p$ is the spin-down timescale. The temporal index $l$ is determined by the braking index and it is often assumed to be $l=2$, i.e., the magnetar spin-down is dominated by dipole radiation. A fraction of this spin-down energy can be thermalized to power the SN luminosity. The idea that pulsars may be able to power SN explosions and SN luminosity appeared shortly after the discovery of pulsars \citep{ostriker1971}. However, the pulsar-powering mechanism for typical core-collapse SNe was not found to match their observations. Later, \citet{maeda2007} proposed that the pulsar power can be an extra energy source to illuminate SNe to explain a peculiar light-curve behavior of SN~2005bf \citep{anupama2005,folatelli2006}. \citet{kasen2010,woosley2010} applied this idea of powering SNe with pulsars for SLSNe. They demonstrated that if the pulsar has an initial rotational period of $\sim 1~\mathrm{ms}$ with a dipole magnetic field of $\sim 10^{14}~\mathrm{G}$, it can reproduce the light curves of Type~I SLSNe. Such a strongly-magnetized, rapidly-rotating pulsar powering SLSNe are often referred to as ``magnetars.''

This magnetar scenario to power SLSN luminosity is currently the most popular scenario to explain SLSNe without clear CSM interaction signatures. Many studies applied the magnetar model to fit the light-curve evolution of Type~I SLSNe to constrain the properties of magnetars powering Type~I SLSNe \citep[e.g.,][]{inserra2013mag,metzger2015diversity,kashiyama2016}. The magnetar model can explain early-phase light curves of both slowly-evolving and rapidly-evolving Type~I SLSNe well. The systematic model fitting to the magnetar-powered model for the Type~I SLSN sample from ZTF obtained the average initial spin period of $2.65^{+2.58}_{-0.68}~\mathrm{ms}$ and the dipole magnetic field strength of $0.98^{+0.98}_{-0.63}\times 10^{14}~\mathrm{G}$ with the $1\sigma$ range \citep{chen2023ztf2}. This fitting provides the average ejecta mass of $5.03^{+4.01}_{-2.39}~\mathrm{M_\odot}$ and the average kinetic energy of $2.13^{+1.89}_{-0.96}\times 10^{51}~\mathrm{erg}$ with the $1\sigma$ range \citep{chen2023ztf2}. The distributions of the estimated parameters and their correlation can be found in Figure~\ref{chapslsn:magnetar}. Under the assumption of the magnetar spin-down model, a possible correlation between SN ejecta mass and initial spin periods is found \citep{blanchard2020}.

The observed complexity in the light curves of Type~I SLSNe is also suggested to be explained by the magnetar scenario. The precursor bump can be explained by the additional shock breakout of a strong shock formed by the central energy injection at the center \citep{kasen2016,liu2021}. The late-phase light curves powered by magnetar spin down are strongly affected by the uncertain thermalization efficiencies of the spin-down energy \citep{kotera2013,wang2015}. In order to distinguish magnetar spin-down and other energy source such as the $^{56}$Ni radioactive decay, light curves need to be followed for around 1000~days \citep{moriya2017magni}, but SLSNe with such a long-term observation is still limited \citep{blanchard2017}. The late-phase light curve undulations can be related to temporal changes in magnetar activities \citep[e.g.,][]{yu2017flare,moriya2022murase}.

Spectroscopic properties are also found to be consistent with the magnetar model \citep[e.g.,][]{jerkstrand2017,dessart2019magmod}. Especially, the brightness in ultraviolet in early phases as well as slow evolution in the photospheric velocity is consistent with the predicted properties from the magnetar powered models. Aspherical nature of Type~I SLSNe observed in some Type~I SLSNe can also be explained by the magnetar scenario because the spin-down energy injection inevitably occur in an aspherical form. Jet emergence from magnetars is also expected \citep[e.g.,][]{soker2022}.

The magnetar energy injection at the central region of SN ejecta is predicted to result in several interesting characteristic features in SLSNe. For example, the extra strong energy input at the innermost layers of the SN ejecta can lead to strong mixing within the ejecta and make the ejecta density structure flatter than other SNe \citep[e.g.,][]{chen2016k,suzuki2017mag,desai2023}. Such a flat density structure may affect spectral formation in SLSNe, although their consequences have not been studied in detail. Another characteristic prediction is that high energy and X-ray emissions can be observed at late phases after the central nebular regions formed by the spin down of the central magnetar become transparent \citep[e.g.,][]{metzger2014ionbreakout,murase2015}. In addition, magnetar-powered SLSNe can become bright in radio because of the pulsar wind nebula formed by the central magnetar. However, X-ray and radio emissions from Type~I SLSNe are often not consistent with simple predictions \citep[e.g.,][]{margutti2018,murase2021}. The thermalization process in the magnetar spin down is not understood well and more investigations are required to predict their expected observational properties \citep{margalit2018,vurm2021}. Possible gamma-ray emission from SN~2017egm is suggested to be consistent with the magnetar spin-down model \citep{li2024gamma}. If the central magnetar is massive enough to require rotational energy to sustain its mass, the magnetar would eventually collapse to a black hole at some moment and leave some observational consequences \citep{metzger2015diversity,moriya2016magbh}. Finally, FRBs might be associated with SLSNe if they are both powered by magnetars \citep{metzger2017frb}, although host galaxy properties of SLSNe and FRBs are different (Section~\ref{sec:host}).

The progenitors of magnetar-powered SLSNe should have rapid rotation. Rapidly rotating Type~I SLSN progenitors could be realized through chemically homogeneous evolution \citep{aguilera-dena2018,aguilera-dena2020} possibly in binary systems. The strong magnetic field may exist at the time of core collapse or amplified during the core collapse. Stellar mergers may also be responsible for the magnetic field amplification \citep{ablimit2022}. We note that some Type~I SLSNe showing late-phase CSM interaction features require an additional CSM component ejected from the progenitor in addition to the rapid rotation and strong magnetic field. Some Type~I SLSNe show late-phase hydrogen-rich emission that requires hydrogen-rich mass loss from the progenitors shortly before their explosions. 

\begin{figure}[t]
\centering
\includegraphics[width=.7\textwidth]{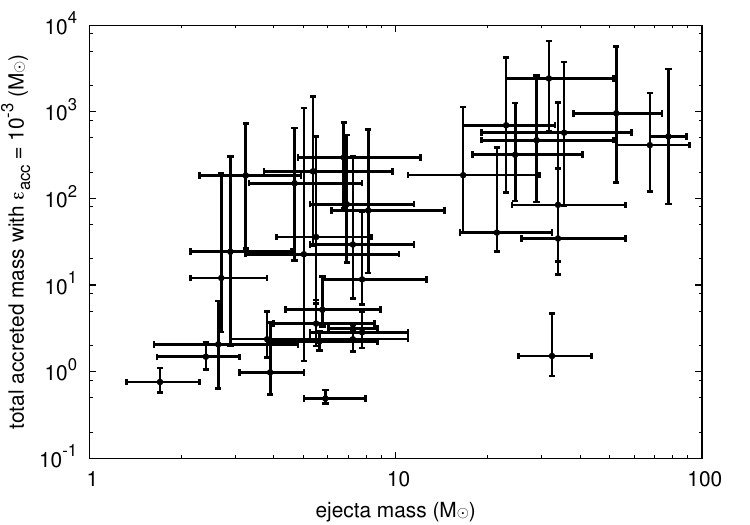}
\caption{Ejecta mass and total accreted mass required to account for Type~I SLSN light curves in the fallback accretion model. The energy conversion efficiency of $\varepsilon_\mathrm{acc}=10^{-3}$ is assumed. Modified from \citet{moriya2018fallback}.}
\label{chapslsn:fallback}
\end{figure}

\subsection{Black hole accretion}
A black hole, instead of a neutron star, can be formed after the terminal collapse of a massive star. Accretion towards the black hole can launch a jet or a disk wind outflow that can be a luminosity source of SNe. If a massive star collapses directly to the black hole with a free-fall timescale, the accretion timescale would be too short to power SLSNe. However, if the outer layers of the SN ejecta are first ejected and then fall back onto the black hole, a long-lasting mass accretion towards the central black hole can be realized. Such a fallback accretion is suggested to be a potential power source for SLSNe \citep{dexter2013}. In a simplified picture, the luminosity input from the accretion can be expressed as
\begin{align}\label{eq:acc}
L_\mathrm{accretion}=\varepsilon_\mathrm{acc}\dot{M}_\mathrm{acc} c^2,
\end{align}
where $\varepsilon_\mathrm{acc}$ is the thermalization efficiency of the accretion and $\dot{M}_\mathrm{acc}$ is the accretion rate to the black hole. The thermalization efficiency is quite uncertain, but it can be reasonably assumed to be $\varepsilon_\mathrm{acc}\sim 10^{-3}$ \citep{dexter2013}. When the fallback accretion is dominant, $\dot{M}_\mathrm{acc}$ becomes proportional to $t^{-3/5}$. \citet{moriya2018fallback} systematically investigated the properties of the fallback accretion required to account for Type~I SLSNe. Although their light curves can be fitted well by the fallback accretion scenario, the required mass to accrete onto the black hole is often found to be very massive (Figure~\ref{chapslsn:fallback}).

\subsection{Other proposed mechanisms}
It is still possible that some unrecognized mechanisms lead to at least some fraction of the observed SLSN populations. For example, axion-instability SNe might be related to SLSNe \citep{sakstein2022,mori2023}. A latent energy released by the phase transition from neutron stars to quark stars is suggested to power SLSNe \citep[e.g.,][]{2012MNRAS.423.1652O}. Other unconsidered energy sources may also play an important role.

\section{Outlook}
As we face the era of large-scale transient surveys in various wavelengths, the number of SLSN discovery is still expected to increase. For example, the Vera C. Rubin Observatory's Legacy Survey of Space and Time (LSST) is expected to discover $\sim 10^4$ SLSNe in a year, although only a limited fraction among them will have sufficient light-curve information to reconstruct their physical properties \citep{villar2018,dubuisson2020}.

Unexplored frontier in SLSN discoveries are at high redshifts. Only a couple of SLSNe have been observed at $z\gtrsim 2$ so far \citep{cooke2012,pan2017,smith2018,moriya2019hsc,curtin2019}. SLSNe at high redshifts will allow us to explore SLSN properties in, e.g., low metallicity environments that will be essential information to uncover their progenitors. In addition, the event rates of PISNe are predicted to be higher at higher redshifts \citep{briel2022} and we expect to discover many PISNe if we search for them at high redshifts. Exploring high-redshift SLSNe require near-infrared transient surveys. Fortunately, several wide-field, sensitive near-infrared imaging instruments will be available in the coming years and they will allow us to explore SLSNe at high redshifts \citep{tanaka2012,tanaka2013}. For example, Euclid has started its operation with successful SN discoveries \citep{euclid2024}. Euclid is expected to discover dozens of SLSNe and PISNe up to $z\sim 4$ \citep{inserra2018,moriya2022euclid}. Nancy Grace Roman Space Telescope, which is currently planned to be launched in 2026, can realize time domain surveys that allow us to discover SLSNe beyond $z\sim 6$ \citep{moriya2022roman}. While the field-of-view is small, James Webb Space Telescope may also potentially discover high-redshift SLSNe \citep{regos2019,venditti2024}. SLSNe at high redshifts may be used as distance measurement \citep[e.g.,][]{inserra2021cos,wei2015cosmology,khetan2023nandita} as well as light sources to explore interstellar media in the distant galaxies that hosted SLSNe \citep[e.g.,][]{berger2012}.

\begin{ack}[Acknowledgments]

TJM is supported by the Grants-in-Aid for Scientific Research of the Japan Society for the Promotion of Science (JP24K00682, JP24H01824, JP21H04997, JP24H00002, JP24H00027, JP24K00668) and the Australian Research Council (ARC) through the ARC's Discovery Projects funding scheme (project DP240101786).
\end{ack}

\seealso{\citet{gal-yam2012,gal-yam2019review,howell2017,moriya2018,inserra2019,nicholl2021review}}

\bibliographystyle{Harvard}
\bibliography{reference}

\end{document}